\documentclass[11pt]{article}       
 \usepackage{amsmath}
\usepackage{graphicx}
\usepackage{amsfonts}
\usepackage{authblk}

\newcommand{\p}{\mathcal{P}}

\begin{document}

\title{{\em rasbhari}: Optimizing Spaced Seeds
for Database Searching, Read Mapping and
Alignment-Free Sequence Comparison}

\author[1]{Lars Hahn} 
\author[1]{Chris-Andr\'{e} Leimeister}
\author[2]{Rachid Ounit} 
\author[2]{Stefano Lonardi} 
\author[1,3]{Burkhard Morgenstern}

\affil[1]{ University of G\"ottingen, Department of Bioinformatics, 
Goldschmidtstr. 1, 37077 G\"ottingen, Germany}  

\affil[2]{University of California, Riverside, Department of Computer Science
and Engineering, 900 University Ave.  Riverside, CA 92521, USA}
 
\affil[3]{University of G\"ottingen, Center for Computational Sciences,  
Goldschmidtstr. 7, 37077 G\"ottingen, Germany}


\maketitle 

\begin{abstract}

Many algorithms for sequence analysis rely on word matching or word statistics.
 Often, these approaches can be improved if
binary  patterns representing {\em match} and {\em don't-care}
positions are used as a filter, 
such that only those positions of words are considered that correspond 
to the {\em match positions} of the patterns.
The performance of these approaches, however,  
depends on the underlying  patterns.  
Herein, we show that the {\em overlap complexity} of a pattern set
that was introduced by Ilie and Ilie is closely
related to the {\em variance} of the number of 
matches between two evolutionarily related sequences with respect to this  
pattern set. 
We propose a modified hill-climbing algorithm to optimize
pattern sets for database searching,
read mapping and alignment-free sequence comparison of nucleic-acid sequences;
our implementation of this algorithm is called {\em rasbhari}.
Depending on the application at hand, {\em rasbhari} can either minimize the
{\em overlap complexity} of pattern sets, maximize their {\em sensitivity}
in database searching 
or minimize the {\em variance} of the number of pattern-based matches in  
alignment-free sequence comparison.  
We show that, for database searching, {\em rasbhari} 
generates pattern sets with slightly
higher {\em sensitivity} than existing approaches. In our {\em Spaced Words}
approach to alignment-free sequence comparison,
pattern sets calculated with {\em rasbhari} led
to more accurate estimates of phylogenetic distances
than the randomly generated pattern sets that we previously used. 
Finally, we used {\em rasbhari} to generate patterns for short 
read classification with {\em CLARK-S}. Here too, the sensitivity 
of the results could
be improved, compared to the default patterns of the program.  
We integrated {\em rasbhari} into {\em Spaced Words};  
the source code of {\em rasbhari} is freely available at  
{\tt http://rasbhari.gobics.de/}

\end{abstract}

\section*{Author Summary}
We propose a fast algorithm to generate spaced seeds for
database searching, read mapping and alignment-free sequence
comparison. Spaced seeds -- {\em i.e.} patterns of match and 
don't-care positions -- are used by many algorithms for sequence
analysis; designing optimal seeds is therefore an active field 
of research.  
In sequence-database searching, one wants to optimize sensitivity,
{\em i.e.} the probability of finding a region of homology; 
this can be done by minimizing the so-called 
overlap complexity of pattern sets.  In alignment-free DNA 
sequence comparison, the number $N$ of pattern-based matches 
is used to estimate phylogenetic distances. 
Here, one wants to minimize the variance of $N$ 
in order to obtain stable phylogenies. 
We show that for
spaced seeds, the overlap complexity -- and therefore the sensitivity 
in database searching -- is closely related to the variance of $N$. 
Our algorithm can optimize the sensitivity, overlap complexity
or the variance of $N$, depending on the application at hand.

\section*{Introduction}
\label{intro}
$k$-mers, {\em i.e.} words of length~$k$, are used in  
many basic algorithms for biological sequence comparison.  
Word matches are used, for example, as {\em seeds} in the {\em hit-and-extend}
approach to database searching and read mapping  
\cite{alt:etal:90,sch:mar:zyt:etal:12,hau:sin:rei:14}.
In alignment-free
sequence comparison, sequences are represented as word-frequency 
vectors to estimate distances or similarities between them, 
{\em e.g.} as a basis for phylogeny reconstruction
\cite{cho:hor:lev:09,sim:jun:wu:kim:09,vin:car:fra:etal:12,jun:sim:wu:kim:10,
all:rho:sul:16},
see \cite{vin:14,ber:cha:rag:16} for reviews.
Similarly, word statistics are used to classify DNA or protein 
sequences 
\cite{les:esk:coh:wes:nob:04,oun:wan:clo:lon:15,mei:15}, for datamining \cite{mei:tec:mor:mer:04}
 and for remote homology detection \cite{lin:mei:06}.    
It is well known that many word-based approaches produce better results 
if {\em spaced words} or {\em seeds}  
are used instead of the previously used {\em contiguous} words or word matches. 
That is, for a pre-defined binary 
pattern~$P$ representing {\em match} and {\em don't-care}
positions, one considers only those positions in a word of 
the same length that correspond to the {\em match} positions of~$P$.

Pattern-based word matching  
has been proposed for {\em hit-and-extend} data\-base searching 
by Ma {\em et al.}~\cite{ma:tro:li:02},
see also \cite{cal:rig:93b};
{\em multiple} spaced seeds are now a standard {\em filtering} step in 
homology searching \cite{fri:noe:14,buc:xie:hus:15}.
Spaced seeds are also routinely used  
in metagenome sequence clustering and classification 
\cite{bri:syk:kuc:15,oun:lon:15},  
protein classification \cite{ono:shi:13}, read mapping 
\cite{rum:lac:dal:etal,noe:gir:kuc:10}, to find {\em anchor points}  for
multiple sequence alignment  
\cite{dar:tre:zha:etal:06,dar:mau:per:10}
and for alignment-free phylogeny reconstruction \cite{lei:bod:hor:lin:mor:14};
similarly, the {\em average common substring} approach to sequence
comparison \cite{uli:bur:tul:cho:06} could be improved by allowing for mismatches 
\cite{hau:pie:moe:wie:05,lei:mor:14,tha:cho:liu:etal:15,tha:cho:liu:apo:alu:15,tha:apo:alu:16}. 
Brejova {\em et al.} extended the concept of {\em spaced seeds} to 
homologies among protein-coding regions \cite{bre:bro:vin:04a} and
introduced  
{\em vector seeds} \cite{bre:bro:vin:05}. 
In general, the advantage of pattern-based approaches 
is the fact that spaced-word occurrences at 
neighbouring sequence positions are statistically less dependent
than occurrences of contiguous words \cite{li:ma:zha:06,mor:zhu:hor:lei:15}.

In pattern-based approaches,  the underlying
patterns of match and don't-care positions are of crucial importance 
for the quality of the results. Generally, non-periodic patterns are 
preferred since they minimize redundancies between overlapping matches
and lead to a more even distribution of matches. 
This increases the probability of obtaining a hit between
two homologous sequences in database searching and leads to more stable
distance estimates in phylogeny reconstruction.  
No{\'e} and Martin \cite{noe:mar:14} defined a {\em coverage criterion}
for multiple spaced seeds and showed that this criterion is related to the
{\em Hamming distance} between two sequences. 
 In the context of database searching, patterns or sets of patterns are often called {\em seeds}. (Originally, the word {\em seed} denoted a match of 
-- contiguous or spaced -- words between a query and a database sequence that
could be extended to the left and to the right. But now {\em seed} often
denotes the underlying pattern in pattern-based approaches).  

In database searching, one wants to 
maximize the {\em sensitivity} of pattern sets 
{\em i.e.} the probability of finding at least one hit within a gap-free 
alignment of a given length $L$ and probability $p$ for a match between 
two residues.  
Calculating the sensitivity of a 
pattern set is {\em NP-hard} \cite{li:ma:zha:06}.  The sensitivity can be approximated  
by dynamic programming \cite{ma:tro:li:02,li:ma:kis:tro:04}, but the run time
of this algorithm is still exponential in the length of the pattern.
In {\em PatternHunter II}, a {\em greedy} algorithm  is used to find
suitable patterns. 
In 2007, 
Ilie and Ilie introduced the {\em overlap complexity} of a pattern set
and showed experimentally that -- for a given number of patterns with a 
given  length and number of match positions --
minimizing the overlap complexity corresponds to maximizing 
the {\em sensitivity} in database searching \cite{ili:ili:07}. 
In contrast to the sensitivity, however, the overlap complexity 
can be easily calculated. To find optimal pattern sets, Ilie and Ilie proposed 
a {\em hill-climbing algorithm} that minimizes the overlap complexity.   
They implemented their algorithm in a software tool called {\em SpEED} 
\cite{ili:ili:big:11}, which is several orders of magnitude faster than
competing approaches and is now considered the state-of-the-art
in seed optimization.  

Recently, we proposed to use {\em spaced-word} frequencies instead of 
word frequencies for alignment-free sequence comparison 
\cite{lei:bod:hor:lin:mor:14,hor:lin:bod:etal:14}. We showed  that 
phylogenetic trees calculated from spaced-word frequencies are
more accurate than trees calculated from contiguous-word frequencies. 
As in database searching, our results could be improved by using 
{\em multiple} patterns.
In our original study, we used randomly generated multiple patterns 
of {\em match} and {\em don't-care} positions. 
In a follow-up paper, we studied the 
number~$N$ of spaced-word matches between
two DNA sequences for a set of binary patterns 
\cite{mor:zhu:hor:lei:15}. Our data suggest that minimizing the
variance of $N$ for pattern sets improves 
alignment-free phylogeny reconstruction.  

In this paper, we first show that the variance of the number~$N$
of spaced-word matches is  
closely related to the {\em overlap complexity} of the underlying set
of patterns.  
We propose a modified hill-climbing algorithm that can be used to generate
pattern sets, either with minimal  variance of $N$, or with minimal 
overlap complexity, or with maximal sensitivity in database searching,
depending on the application at hand.
While the algorithm proposed in \cite{ili:ili:07} iterates over all patterns~$P$ in a set ${\mathcal P}$ of patterns and all pairs of positions in $P$ to improve ${\mathcal P}$, we calculate for each pattern $P\in\p$ how much $P$ contributes to the variance or overlap complexity, respectively, of $\p$.  We then modify those patterns first that  contribute most to the variance or complexity.  

The implementation of our approach is called 
{\em rasbhari (Rapid Approach for Seed optimization Based on a 
Hill-climbing Algorithm that is Repeated Iteratively)}.  
Experimental results show that pattern sets calculated with {\em rasbhari}  
have a slightly higher sensitivity in database searching 
than pattern sets calculated 
with {\em SpEED}, while the run time of both programs is comparable. 
In alignment-free sequence comparison,  
we obtain more accurate phylogenetic distances  
if we use {\em rasbhari} to minimize the variance of $N$
for the underlying pattern sets, than we obtained with the randomly generated
pattern sets that we previously used.  
In a third application,  we used pattern sets generated with {\em rasbhari}
in the program {\em CLARK-S} \cite{oun:lon:15} for short read 
classification. 
The sensitivity of the classification could be  
improved in this way, while {\em rasbhari} is substantially
faster than the method that is used by default for pattern generation  
in {\em CLARK-S}.

\section*{Methods}
\subsection*{Overlap complexity} 
We consider sets ${\mathcal P} = \{P_1,\dots,P_m\}$  of binary patterns, 
where $\ell_r$ is the length of pattern $P_r$  
and $\ell = \max_r\ell_r$.  
That is, each $P_r$ is a word of length $\ell_r$ over the alphabet $\{1,0\}$. 
 A `1' in a pattern $P_r$ represents a {\em match} position, a `$0$' 
a {\em don't-care} position. For a single pattern~$P_r$, 
the number of match positions is called its {\em weight}~$w$.
For simplicity, we assume that all patterns in a set ${\mathcal P}$ have the same
{\em weight}. 

In \cite{mor:zhu:hor:lei:15}, we considered for two patterns $P_r, P_{r'}$ 
 and $s \in \mathbb{Z}$ 
the number $n(P_r,P_{r'},s)$ of positions 
that are match positions of $P_r$ {\em or} match positions of $P_{r'}$
(or both), if $P_{r'}$ is  shifted by $s$ positions to the right, 
relative to $P_r$.
If $s$ is negative, $P_{r'}$ is shifted to the left. 
For $P_r = 101011, P_{r'} = 111001$, for example, if $P_{r'}$ is  shifted by
2 positions to the right, relative to $P_r$, then  
there are 6 positions (marked by asterisks below) that are match positions 
of $P_r$ or $P_{r'}$. Thus, for $s=2$, we have  $n(P,P_{r'},2) = 6$: 

\[
\begin{array}{cccccccccc}
P_r:    &   & 1 & 0 & 1 & 0 & 1 & 1 &   &  \\
P_{r'}: &   &   &   & 1 & 1 & 1 & 0 & 0 & 1\\
        &   & * &   & * & * & * & * &   & *\\
        &   &   &   & \$&   & \$&   &   &  \\
\end{array}
\]

For the same situation, Ilie and Ilie
 \cite{ili:ili:07} 
 defined $\sigma[s] = \sigma_{r,r'}[s]$ as the number of positions
where $P_r$ {\em and}  $P_{r'}$ have a match positions, 
such as the positions marked by '\$' above. 
In the above example one would therefore have $\sigma[2] = 2$.  
The {\em overlap complexity (OC)} of a set of patterns $\p=\{P_1,\dots,P_m\}$ 
is then defined in \cite{ili:ili:07} as 
\begin{eqnarray} 
\label{eq_oc} 
  && \sum_{r \le r'} \sum_{s=1-\ell_{r'}}^{\ell_{r}-1} 2^{\sigma_{r,r'}[s]} 
\end{eqnarray}
Note that, since for any two patterns $P_r, P_{r'}$ and $s\in\mathbb{Z}$, the equality 
$$\sigma_{r,r'}[s] = 2  w - n(P_r,P_{r'},s)$$
holds, the {\em overlap complexity} of a set $\p$ can be written as 
\begin{eqnarray}
\label{oc_alt}
 \sum_{r \le r'} 
\sum_{s=1-\ell_{r'}}^{\ell_{r}-1}
2^{\sigma_{r,r'}[s]}  
&=&
2^{2 w}\cdot \sum_{r \le r'} \sum_{s=1-\ell_{r'}}^{\ell_{r}-1} (1/2)^{n(P_r,P_{r'},s)}
\end{eqnarray}
Consequently, if we are looking at sets $\p$ of $m$ patterns with fixed weight $w$ and lengths $\ell_r$, then   minimizing the overlap complexity of ${\mathcal P}$ is equivalent to  minimizing the sum \begin{eqnarray} \label{olc} &&\sum_{r \le r'} \sum_{s=1-\ell_{r'}}^{\ell_{r}-1} (1/2)^{n(P_r,P_{r'},s)}  \end{eqnarray}

Ilie and Ilie showed experimentally that the {\em OC}  is closely related 
to the {\em sensitivity} of a pattern set. More precisely, they showed 
that for pattern sets with  a given number of patterns of given lengths 
and weight, 
minimizing the {\em OC} practically amounts to maximizing the
{\em sensitivity}. 
Consequently, in order to find suitable pattern sets 
for hit-and-extend approaches in database searching, 
they proposed to search for pattern sets with minimal {\em OC}. 
The main advantage of this approach is the fact that  
the {\em OC} of a pattern set  is much easier to calculate than its  
{\em sensitivity}.

\subsection*{Variance of the number of  spaced-word matches}
For a pattern $P$ of length $\ell$, 
we say that two sequences $S_1$ and $S_2$ have a {\em spaced-word match} 
with respect to $P$ at $(i,j)$, if the 
$\ell$-mers starting at $i$ and $j$ have identical characters at all 
{\em match} positions  of~$P$, {\em i.e.} if one has 
$S_1(i+\pi-1) = S_2(j+\pi-1)$ for all match positions $\pi$ in $P$. 
The sequences 
below, for example, have a spaced-word match at $(2,4)$ 
with respect to the {\em pattern} 
$P=110101$. 
Indeed, the $6$-mers starting at 
positions 2 and 4 of the sequences are identical at all positions
corresponding to a {\em match position} (`1') in~$P$, while
positions at {\em don't-care positions} (`0') may be 
matches or mismatches. 
\[
\begin{array}{lccccccccccc}
S_1: &   &   & A & A & T & C & G & A & T & C & A \\
S_2: & C & G & T & A & T & T & G & A & T & T &   \\
$P$: &   &   &   & 1 & 1 & 0 & 1 & 0 & 1 &   &   \\
\end{array}
\]



In \cite{mor:zhu:hor:lei:15}, we considered 
spaced-word matches between two sequences $S_1$ and $S_2$ with respect to a set 
$\p=\{P_1,\dots,P_m\}$ of patterns, so-called $\p$-matches. 
Note that  there can be up to $m$ $\p$-matches at each pair  
 $(i,j)$ of positions of $S_1$ and $S_2$, one $\p$-match for each pattern $P_r$ in $\p$.  
We studied the number $N=N(S_1,S_2,{\mathcal P})$
of ${\mathcal P}$-matches between sequences $S_1$ and $S_2$ 
under a simplified model
of evolution without insertions and deletions, with a {\em match probability} 
$p$ for pairs of homologous positions and a  
{\em background} match probability of~$q$.
It is easy to see that, for a pattern set $\p$, 
the {\em expected} number of $\p$-matches 
depends only on the number $m$ of patterns in $\p$ and on their 
lengths~$\ell_i$ and their weight $w$, {\em i.e.} number of match positions,
but not on the particular sequence of
{\em match} and {\em don't-care} positions in $\p$.  
The {\em variance}  of $N$, however, does depend on the  sequence of 
{\em match} and {\em don't-care} positions.

As discussed in \cite{mor:zhu:hor:lei:15}, many
alignment-free distance or similarity measures are  -- explicitly 
or implicitly -- a function of the number $N$ of (spaced) word matches. 
To obtain stable distance measures for phylogeny reconstruction, 
it is therefore desirable to use pattern sets with a low variance of $N$.
For a given set $\p = \{P_1, \dots, P_m\}$ of patterns of lengths 
$\ell_1, \dots , \ell_m$ and weight $w$, and with the above  simple 
model of evolution, the variance of $N$  can be approximated by   
\begin{equation}
\label{eq_vari}
\begin{split}
Var(N) \approx & \    
 (L-\ell+1) \cdot 
\sum_{r\le r'}\sum_{s\in R(r,r')}  \left(
p^{n(P_r,P_{r'},s)} -  p^{2w} \right)  \\
 & +\ \  (L-\ell+1) \cdot (L-\ell) \cdot 
\sum_{r\le r'}  \sum_{s\in R(r,r')}  \left(
  q^{n(P_r,P_{r'},s)} - q^{2w}\right)  \\  
\end{split}
\end{equation}
where $L$ is the length of $S_1$ and $S_2$, respectively, and 
\[ R(r,r') = \left\{ 
\begin{array}{ll}
\{1-\ell_{r'},\dots,\ell_r-1\} & \text{ if } r < r' \\ 
\{0,\dots,\ell_r-1\} & \text{ if } r = r' \\ 
\end{array}
\right.
\]
is the {\em range} in which $P_{r'}$ is to be shifted 
against $P_r$ \cite{mor:zhu:hor:lei:15}.  Note that
for different patterns $P_{r'}\not= P_{r}$ we have to consider  all 
shifts between 
$1-\ell_{r'}$ and $\ell_r -1$ 
of $P_{r'}$ against $P_{r}$
, for example: 
\[
\begin{array}{rccccccccccccccccccc}
P_r:   &   &   &   &   & 1 & 0 & 1 & 1 &  &         &   & 1 & 0 & 1 & 1 & & & &  \\
P_{r'}:& 1 & 0 & 1 & 0 & 1 &   &   &   & ,&\cdots   & , &   &   &   & 1 &0&1&0&1  \\
s: &&\multicolumn{3}{c}{-4} &&&&&& &&&&&&&3&&\\ 
\end{array}
\]
By contrast,  
if a pattern $P_r$ is shifted against itself, 
only shifts between $0$ and $\ell_r-1$ need to be considered,  
to avoid double counting of shifts\footnote{In  
\cite{mor:zhu:hor:lei:15}, we ignored this fact and gave a slightly
different estimate for $Var(N)$.}, for example: 
\[
\begin{array}{rcccccccccccccc}
P_r: & 1 & 0 & 1 & 1  &   &      &   & 1 & 0 & 1 & 1 & & &  \\
P_r: & 1 & 0 & 1 & 1  & , &\cdots& , &   &   &   & 1 &0&1&1  \\
s:   && \multicolumn{2}{c}{0}&&&&&&&&& \multicolumn{2}{c}{3} & \\ 
\end{array}
\]

\begin{figure}[h]
\begin{center}
\includegraphics[width=6cm]{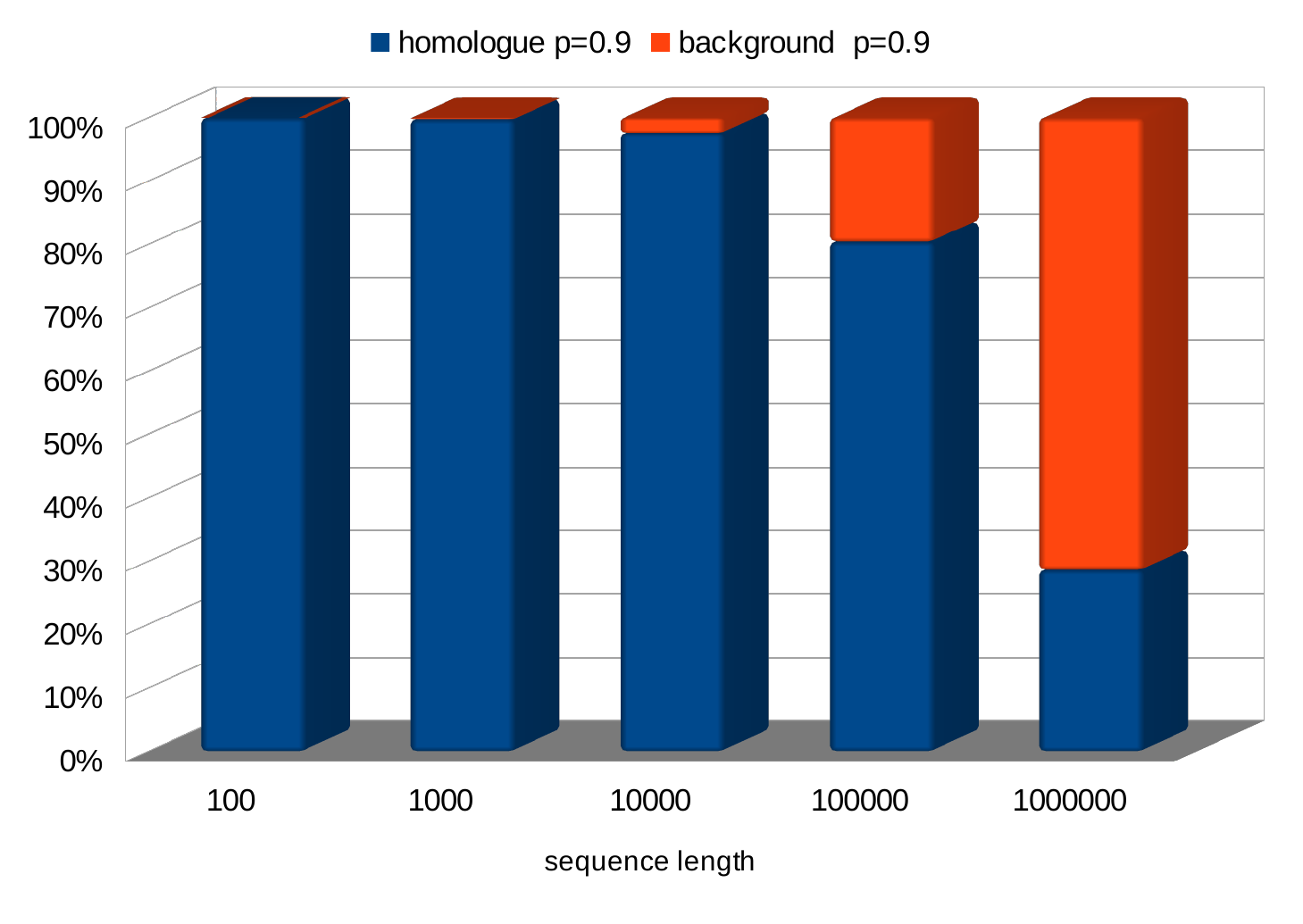}
\includegraphics[width=6cm]{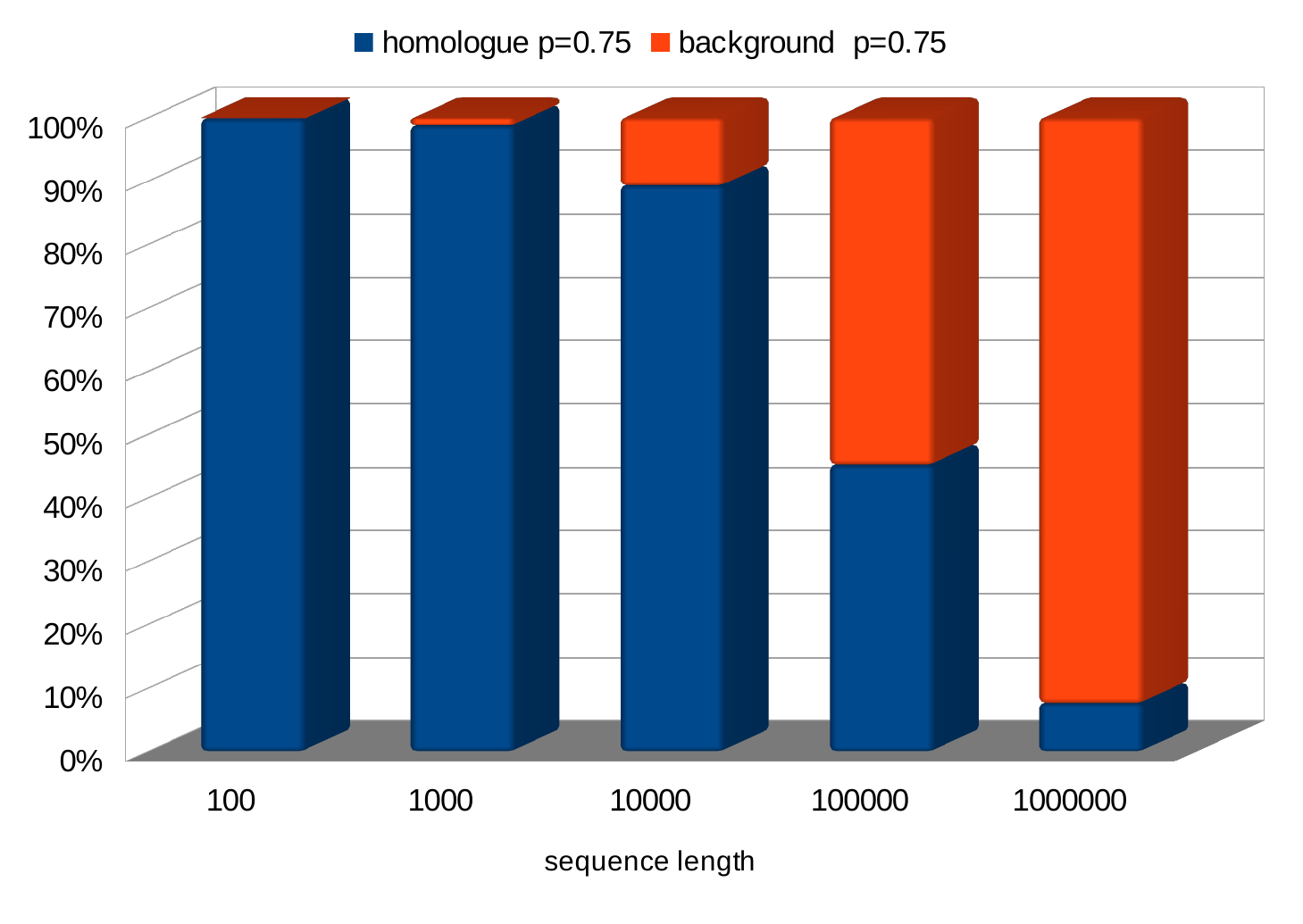}
\includegraphics[width=6cm]{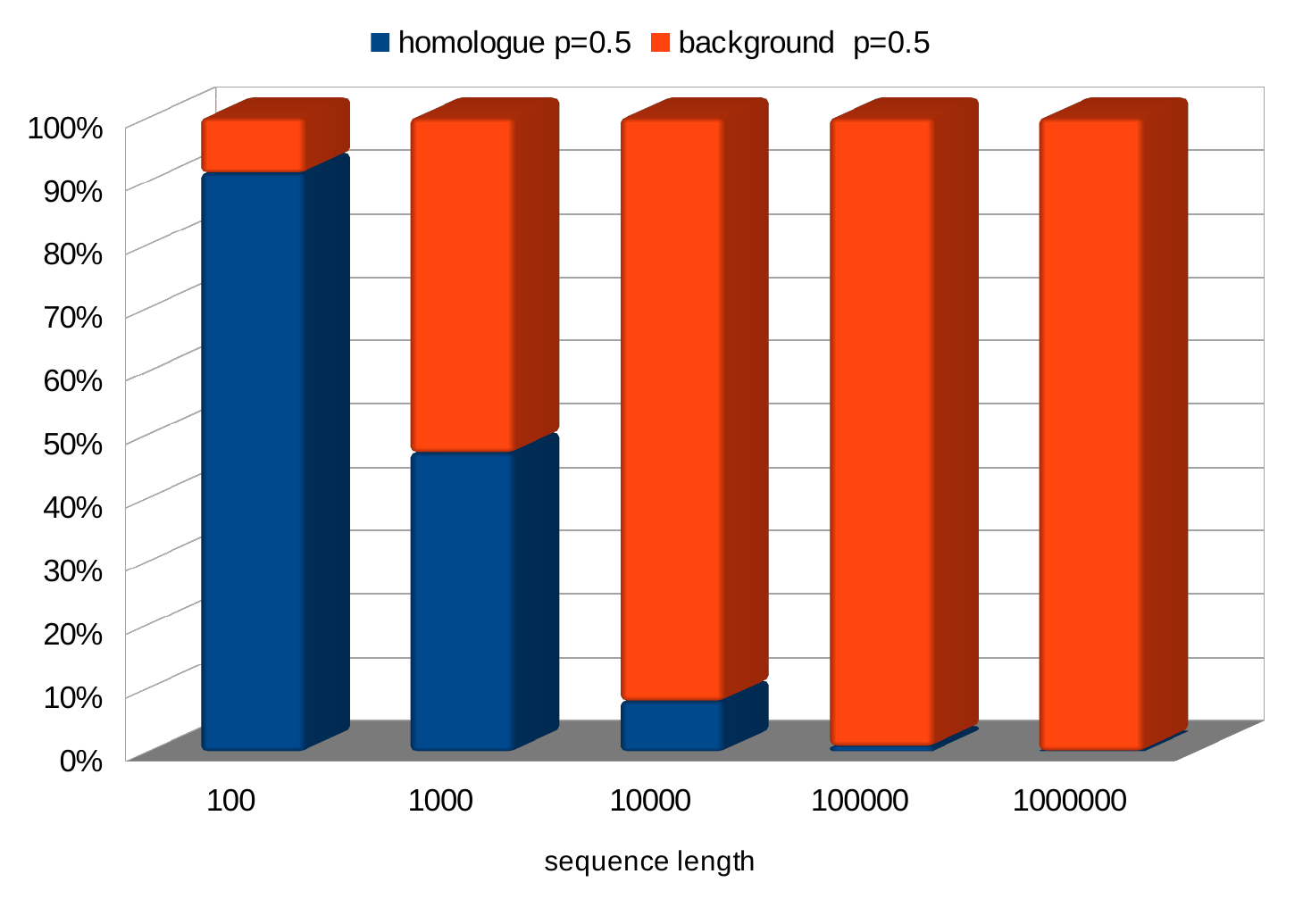}
\end{center}
\caption{Homologue and background variance
Contribution of the {\em homologue} and {\em background} variance
to the total variance of the number
 $N$ of spaced-word matches in equation (\ref{eq_vari})  for different
match probabilities $p$ and sequence lengths $L$.
\label{hom_back}
}
\end{figure}

On the right-hand side of (\ref{eq_vari}), the first summand is the 
variance of the `homologous'
spaced-word matches (in a model without indels, 
these are spaced-word matches involving the same
positions in both sequences), while the second summand comes 
from background matches. The {\em relative} weight of the background
matches in (\ref{eq_vari}) 
depends on the match probability~$p$ and the sequence 
length~$L$;
for $p>>q$ and small $L$, the
variance of $N$ is dominated by the `homologous' term, see
Figure~\ref{hom_back}. 
 Obviously, 
for large~$L$, the background spaced-word matches dominate the `homologous'
ones, since the number of background matches grows quadratically with $L$,
while the  `homologous' matches grow only linearly.

Note that, for $L, \ell$ and $w$ fixed,  minimizing the variance of $N$ 
amounts to minimizing 
\begin{eqnarray}
\label{mini_var} 
&&\sum_{r\le r'} \sum_{s\in R(r,r')} p^{n(P_r,P_{r'},s)} 
\ +\ \  (L-\ell) \cdot  
\sum_{r\le r'} \sum_{s\in R(r,r')} q^{n(P_r,P_{r'},s)} 
\end{eqnarray}
Comparison with (\ref{oc_alt}) shows that, 
in the special case of $p= 1/2$, the first summand of  (\ref{mini_var}) 
that corresponds to the {\em homologous} matches
is almost identical with the {\em overlap complexity} defined by Ilie and Ilie
(except for the range $R(r,r)$ in which a pattern $P_r$ is
shifted against itself). For sequences of moderate length, the
overlap complexity can therefore be seen as an approximation to the variance of the number
of spaced-word matches.

In any case, the overlap complexity and the variance of $N$ for a set of
pattern $\p = \{P_1,\dots,P_m\}$ both have the form 
\begin{eqnarray}
\label{of} 
\sum_{r \le r'}  
\alpha_{r,r'}(\p) &&  
\end{eqnarray}
with 
\begin{eqnarray} 
\label{alpha} 
\begin{split}
\alpha_{r,r'}(\p) = \left\{ 
\begin{array}{ll} 
\displaystyle\sum_{s=1-\ell_{r'}}^{\ell_{r}-1} 2^{\sigma_{r,r'}[s]} & (OC) \\ 
(L-\ell+1)  
\displaystyle\sum_{s\in R(r,r')}
     \left(p^{n(P_r,P_{r'},s)} + (L-\ell)\cdot q^{n(P_r,P_{r'},s)}\right) & (Var) \\  
\end{array}
\right.
\end{split}
\end{eqnarray}
Our optimization problem is therefore: for integers $m, \ell_1, \dots \ell_m, w$,
find a set $\p$ of $m$ patterns of lengths $\ell_1, \dots ,\ell_m$ and weight~$w$ that  
minimizes the sum (\ref{of}).

\subsection*{Hill-climbing algorithms to find sets of patterns with 
minimal $Var(N)$ or {\em OC}} 

Both {\em SpEED} and our new algorithm start with a randomly generated 
pattern set $\p$ and use {\em hill-climbing}  
to gradually reduce the {\em OC} or $Var(N)$ of $\p$. 
After a pattern set with low {\em OC} is obtained in this way, its 
{\em sensitivity} is calculated, if one is looking for a pattern set
with maximal sensitivity. This step is omitted in {\em rasbhari} 
if $Var(N)$  or $OC$  is to be minimized.
The whole procedure is repeated, and the pattern set with the overall 
highest sensitivity -- or lowest variance of $N$ or $OC$,
respectively -- is returned.

\subsubsection*{Original hill-climbing algorithm}

To improve the current pattern set $\p$, 
the hill-climbing algorithm implemented in {\em SpEED} 
looks at all triples $(r,i,j)$ where $P_r$ is a pattern in 
$\p$, and $i$ and $j$ are a {\em match position} and 
 a {\em don't-care} position in  $P_r$, respectively. For each such triple $(r,i,j)$, the algorithm
considers the pattern set that would be obtained from $\p$ by swapping 
$i$ and $j$ in $P_r$  -- {\em i.e.} by turning
$i$ into a don't-care and $j$ into a match position.
The {\em OC}
is calculated for all pattern sets that can be obtained in this way, and the one with the lowest
{\em OC} is selected as the next pattern set $\p$. This is repeated iteratively. 

There are $O(m\cdot \ell^2)$ triples $(r,i,j)$ to be considered to modify the current pattern set $\p$. For each of these triples, the {\em OC} is to be calculated for 
the pattern set that would be obtained by swapping $i$ and $j$ in $P_r$.  
To this end, the modified pattern $P_r$ has to be compared to the $m-1$ remaining patterns in $\p$ which, for each pattern comparison, 
involves $O(\ell)$ shifts of two patterns against each other. In each shift, the number of common match positions is to be counted, which takes again $O(\ell)$ time. 
Thus, calculating the {\em OC} of the pattern set obtained by swapping two positions $i$ and $j$ in a pattern $P_r$ 
takes $O(m\cdot \ell^2)$ time, so finding an optimal triple $(r,i,j)$ to determine the next pattern set 
takes $O(m^2\cdot \ell^4)$ time.
This step is repeated a certain number of times; 
for the pattern set that is finally obtained by this hill-climbing routine, the {\em sensitivity}
is calculated. This whole procedure is repeated 5,000 times, 
and finally the set with the best sensitivity is returned.

\subsubsection*{Modified hill-climbing algorithm}

In our modified hill-climbing algorithm, we also  swap 
a match position $i$ with a don't-care position $j$ 
in some pattern~$P_r$ in each step of the algorithm, and we evaluate
the pattern set that would be obtained by this operation. 
However, instead of looking at {\em all} possible triples $(r,i,j)$, 
we look at those patterns first that contribute most to the {\em OC}
or $Var(N)$, respectively, of the current pattern  set $\p$. 
The contribution 
\begin{eqnarray} 
\label{contrib_oc} 
  C_r & = & \sum_{r'} \alpha_{r,r'}  
\end{eqnarray}
of a pattern $P_r\in\p$ to the {\em OC} or $Var(N)$ of $\p$ can be calculated 
as a by-product, whenever {\em OC} or $Var(N)$ is calculated,
with $\alpha$ as in (\ref{alpha}).  
We then sort the patterns in $P_r\in\p$ according to the values $C_r$,
and we process them in descending order of $C_r$, {\em i.e.} 
we look at those patterns first that contribute 
{\em most} to the {\em OC} or $Var(N)$ of $\p$. 

For the current pattern in the  list, 
we randomly select a match position $i$ and a don't-care position $j$. 
If swapping $i$ and $j$ does {\em  not} improve
the current pattern set, we move on to the next pattern in the list 
and proceed in the same way. 
This is repeated until we find a pattern  where swapping the selected 
pair of random positions does improve $\p$. 
In this case, the modified pattern is accepted, all values $C_r$ are
updated, the patterns in $\p$ are sorted accordingly,  and we start
again with the pattern $P_r$ with maximum $C_r$. 
If we reach the last pattern in the list without obtaining any improvement, 
we start again with the first pattern, {\em i.e.} the pattern with the largest $C_r$,
select new random positions $i$ and $j$ etc. 
Processing one pattern $P_r$ in this way takes $O(m\cdot \ell^2)$ 
time, since we look only at one single pair $(i,j)$ 
and calculate the {\em OC} or $Var(N)$ of the pattern set that would be
obtained by swapping $i$ and $j$ in $P_r$.

The hill climbing is continued until a user-defined
number of triples $(r,i,j)$ have been processed, 
then the current pattern set is returned; 
by default, 25,000 triples are processed.  
If we want to obtain a pattern set with maximal {\em sensitivity}, the
described hill-climbing procedure is repeated 100 times, and for the pattern set
with the lowest {\em OC} among the 100 obtained pattern sets,  
the {\em sensitivity} is calculated. 
To calculate the sensitivity, {\em rasbhari} uses program code from {\em SpEED}.
Again, this whole process is repeated 5,000 times, so for 
a total of 5,000 pattern sets the sensitivity is calculated during one
program run.
This is similar to {\em SpEED}, but in {\em SpEED} the time-consuming sensitivity
calculation is carried out after {\em one} round of hill climbing. By contrast, we run our
faster hill-climbing routine 100 times before we calculate the sensitivity
for the {\em best} pattern set from these 100 runs. 
The final output of our program is the pattern set with the highest
sensitivity from the 5,000 iterations.

The number $m$ of patterns and their weight~$w$ are to be specified by the user.
If $Var(N)$ is to be minimized for alignment-free sequence comparison, 
all patterns have the same length $\ell$ which is also to be specified by
the user. 
If the {\em sensitivity} is to be maximized for database searching and
read alignment, better results are achieved if the patterns in $\p$
have different lengths. In this case, the maximum and minimum
pattern lengths need to be specified. The program then 
selects lengths~$\ell_1, \dots, \ell_m$ that are evenly 
distributed between these extreme values. 

\section*{Results}

\subsection*{Sensitivity in database searching}

\begin{table}[h!]
\caption{Sensitivity of pattern sets  
in hit-and-extend database searching, calculated from different
programs.  
Parameter settings for the number $m$ and weight $w$ of patterns,
the length $H$ of the gap-free homology region between query and database
sequences and the match probability $p$ in the homology regions,
are taken from three popular programs {\em SHRiMP2, PatternHunter II}
and {\em BFAST}. Results of existing programs are taken from their
respective publications.
\vspace{2mm}
\label{sensi_tab}}
\begin{tabular}{|c|c|c|c|c|c|c|c|}
\hline
$w$ & $p$ & {\em Iedera} & {\em SpEED} & {\em AcoSeeD} & {\em FastHC} & {\em MuteHC} & {\em rasbhari} \\
\hline
\multicolumn{8}{|c|}{\textbf{{\em SHRiMP2:} 4 patterns ($H=50$)} }\\
\hline
       & 0.75 & 90.6820& 90.9098  & 90.9513 & 90.7312 & {\bf 92.6812} & {90.9614} \\
10     & 0.80 & 97.7586& 97.8337  & 97.8521 & 97.7625 & {\bf 98.3836} & {97.8554}\\
       & 0.85 & 99.7437& 99.7569  & 99.7614 & 99.7431 & {\bf 99.8356} & {99.7618}\\
\hline
  & 0.75 & 83.2413& 83.3793& \textbf{83.4728} & 83.3068 &83.4127 & 83.4679 \\
11& 0.80 & 94.9350& 94.9861& 95.037           & 94.9453 &95.0194 & \textbf{95.0386}\\
  & 0.85 & 99.2189& 99.2431& 99.2478          & 99.2250 &99.2486 & \textbf{99.2506}\\
\hline
  & 0.80 & 90.3934& 90.5750 & 90.6328 & 90.4735 &90.5820 & \textbf{90.6648} \\
12& 0.85 & 98.0781& 98.1589 & 98.1766 & 98.1199 &98.1670 & \textbf{98.1824}\\
  & 0.90 & 99.8773& 99.8821 & 99.8853 & 99.8771 &99.8836 & \textbf{99.8864} \\
\hline
  & 0.85 & 84.5795& 84.8212 & \textbf{84.9829} &84.6558 &84.8764 &  84.969 \\
16& 0.90 &97.2806 & 97.4321 & 97.4712          &97.3556 &97.4460 &  \textbf{97.5035}\\
  & 0.95 & 99.9331& 99.9388 & 99.9419          &99.9347 &99.9424 &  \textbf{99.9441}\\
\hline
  & 0.85 & 72.1695 & 73.1664 & \textbf{73.27}   &72.9558  & & 73.2209 \\
18& 0.90 & 93.0442 & 93.7120 & 93.7778          &93.6030  & &  \textbf{93.78}\\
  & 0.95 & 99.6690 & 99.7500 & \textbf{99.7599} &99.7399  & & 99.7557\\
\hline
\hline
\multicolumn{8}{|c|}{\textbf{{\em PatternHunterII:} 16 patterns ($H=64$)} }\\
\hline
  & 0.70 & 92.0708& 93.2526& & 93.0585 & &  \textbf{93.4653} \\
11& 0.75 & 98.3391& 98.6882& & 98.6352 & &  \textbf{98.7573}\\
  & 0.80 & 99.8366& 99.8820& & 99.8750 & & \textbf{99.8907}\\
\hline
\hline
\multicolumn{7}{|c|}{\textbf{{\em BFAST:} 10 patterns ($H=50$)} }\\
\hline
  & 0.85 & 60.1535& 60.8127&  & 60.0943 & & \textbf{60.9919} \\
22& 0.90 & 87.9894& 88.5969&  & 88.0426 & & \textbf{88.8005}\\
  & 0.95 & 99.2196& 99.3659&  & 99.2923 & & \textbf{99.4099}\\
\hline
\end{tabular}

\end{table}

To evaluate {\em rasbhari}, we first applied it to generate 
pattern sets,  maximizing  the  {\em sensitivity} 
for database searching and read mapping. 
For the number~$m$ and weight~$w$ of the patterns and for the length~$H$ and 
match probability~$p$
of the homology regions, we used the parameter settings from 
 {\em SHRiMP2}~\cite{dav:dza:lis:ili:bru:11}, 
{\em PatternHunter II}~\cite{li:ma:kis:tro:04} 
and {\em BFAST}~\cite{hom:mer:nel:09}. 
We and compared it to the 
sensitivity of pattern sets  obtained with 
{\em Iedera}~\cite{kuc:noe:roy:06}, {\em SpEED}~\cite{ili:ili:big:11}, 
{\em AcoSeeD}~\cite{duc:din:dan:etal:12},  {\em FastHC} and {\em MuteHC}~\cite{do:tra:15}
as published by the authors of these programs;   
the results of this comparison are shown in Table \ref{sensi_tab}.
Here, the sensitivity values of  {\em AcoSeeD}  are
{\em average} values over 10 program runs reported 
in~\cite{duc:din:dan:etal:12}. 
If the sensitivity of a pattern set is to be optimized, 
the run time of {\em rasbhari} is comparable to {\em SpEED}, since  the 
most time-consuming step in both programs is to calculate the 
sensitivity of a current pattern set~$\p$ which is done 5,000 times per  
program run in each of the two programs.  

\subsection*{Alignment-free phylogeny reconstruction} 

Next, we wanted to know how 
alignment-free phylogeny reconstruction can be improved with  
{\em rasbhari}. To this end, we estimated pairwise distances
between simulated DNA sequences using the {\em Spaced Words}
approach described in \cite{mor:zhu:hor:lei:15}, and we measured the 
accuracy of these distance estimates for different underlying pattern sets. 
Here, we used {\em rasbhari} to minimize the  {\em variance} of the number~$N$
of spaced-word matches between two sequences.  
Since there is no other method to minimize the variance of~$N$, we 
compared the pattern sets from {\em rasbhari} with the randomly generated 
pattern sets that we initially used. 
%
The phylogenetic distances obtained from both program runs were 
compared to the `real' distances, {\em i.e.} the average number of 
substitutions per position in the sequences. 

\begin{table}[h!]
\caption{
Accuracy of pattern sets in alignment-free phylogeny reconstruction. 
Evolutionary
distances between simulated DNA sequences were estimated based on the 
number~$N$ of spaced-word matches between them, using the  
method published in \cite{mor:zhu:hor:lei:15}.
Four sets of sequence pairs were generated, such that for each set 
the substitution rates of the sequence pairs are 
within a certain range, as specified in the table. 
Pattern sets were generated with {\em rasbhari}, minimizing
the variance of~$N$, and compared to the randomly generated pattern sets
that we previously used. 
The table shows the mean quadratic difference between the estimated distances 
and the `real' distances, {\em i.e.} the number of substitutions per position
in the model used to generate the sequence pairs.  
\label{dist_tab}}
\vspace{2mm}
\begin{tabular}{|l|c|c|c|c|}
\hline
Substitutions per position 
& 0 - 0.25  & 0.25 - 0.5 & 0.5 - 0.75 & 0.75 - 1.0 \\ 
\hline 
Random pattern sets   & 1.86E-06 & 5.10E-05  & 1.92E-03   & 3.11E-02 \\
{\em rasbhari}    & 1.69E-06 & 4.51E-05  & 1.60E-03   & 3.01E-02 \\  
\hline
\end{tabular}
\end{table}

As test data, we generated four data sets with 
10,000 pairs of DNA sequences of length 100 $kb$ each. In each of these
data sets, we used a specific range of substitution rates as  
shown in Table  \ref{dist_tab}; the substitution rate for a simulated
 sequence pair was then randomly picked within the respective range. 
For each program run, we used a set of $m=3$ patterns of length 20 
with 16 {\em match}  and 4 {\em don't-care} positions.  
Table \ref{dist_tab} shows the {\em mean quadratic difference} between
the estimated distances and the `real' distance between the sequences.  
In all four categories, the pattern sets  generated
with {\em rasbhari} were superior to the randomly generated pattern sets. 

\subsection*{Read classification}

As a third test case, we used different  pattern sets for 
{\em CLARK-S} \cite{oun:lon:15}, a recently developed tool 
for short read classification. We evaluated the 
accuracy of {\em CLARK-S} with three underlying pattern sets, namely 
{\bf (A)} with the patterns used by default in the program, {\bf (B)}
with patterns from {\em rasbhari} minimizing 
{\em overlap complexity} and {\bf (C)} with patterns from {\em rasbhari} 
maximizing {\em sensitivity}.  
{\em CLARK-S} uses sets of $m=3$ patterns of length $\ell = 31$ and 
with a weight of $w=22$. Since {\em SpEED} is too slow to
generate pattern sets for these parameters, the authors generated
pattern sets for {\em CLARK-S} by exhaustively searching over all {\em single}
patterns with $\ell=31$ and $w=22$. If the first and the last position in 
the patterns are required to be {\em match positions}, this approach has
to evaluate  
${29 \choose 20} \approx 10^7 $ possible patterns.  The 
sensitivity of each of these patterns was calculated, and the three patterns
with the highest sensitivity were selected. 
Note however, that maximizing the sensitivity
of {\em single} patterns is only an approximation
to finding a {\em set} of patterns with maximal {\em total} sensitivity.

\begin{figure}[h!]
\begin{center}
\includegraphics[width=9cm]{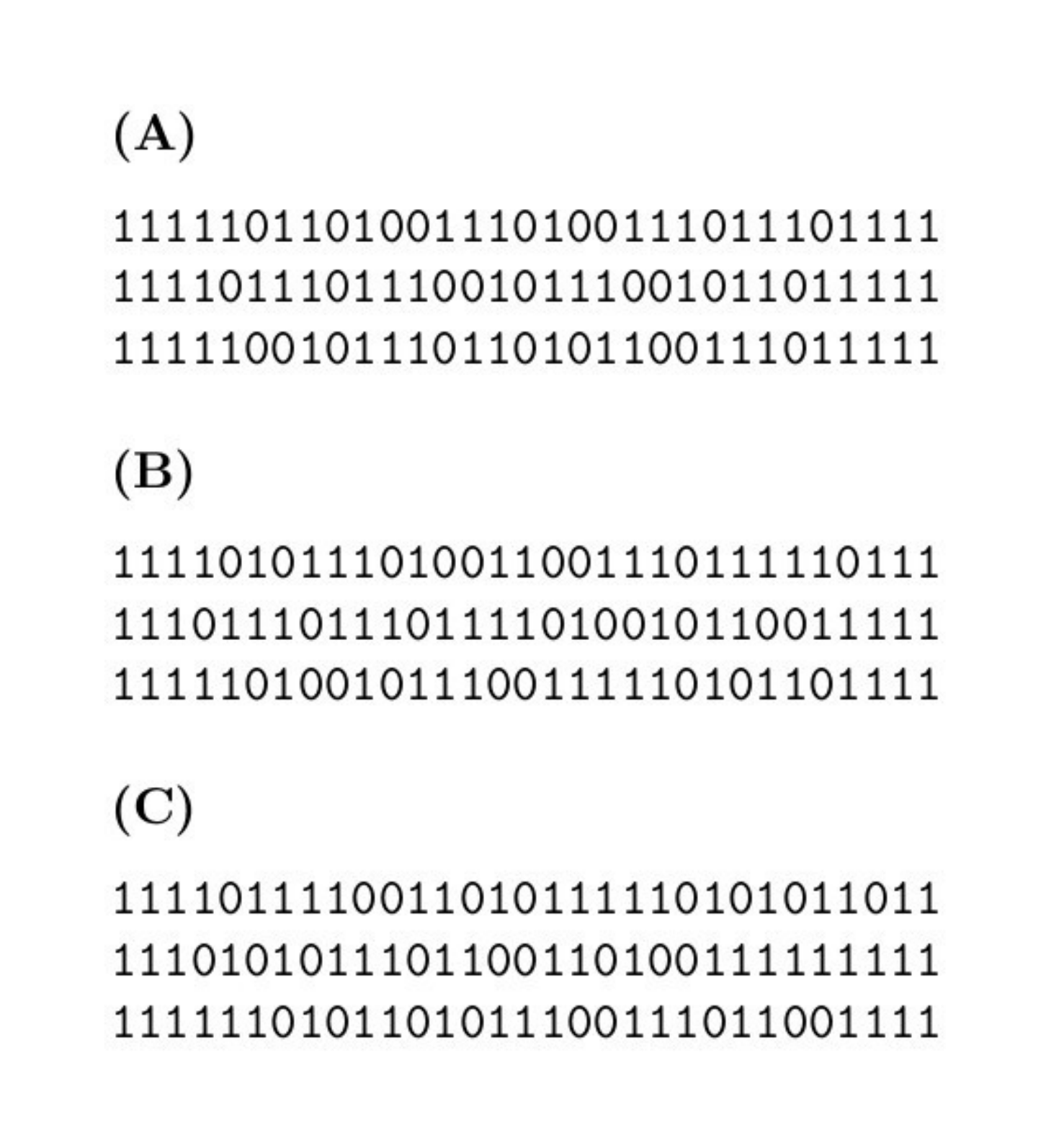}
\end{center}
\caption{Pattern sets used for short read classification: {\bf (A)} 
as used by default in {\em CLARK-S}, {\bf (B)} generated with {\em rasbhari}
minimizing {\em overlap complexity} and {\bf (C)} generated with 
{\em rasbhari} maximizing {\em sensitivity}.   
\label{fig_patterns}
}
\end{figure}

Figure \ref{fig_patterns} shows the default pattern set from {\em CLARK-S}
and the two pattern sets  generated by {\em rasbhari} as described. 
The exhaustive procedure used by {\em CLARK-S} took 2~hours to generate the
pattern set. 
{\em rasbhari}, by contrast, calculated pattern sets with the same
parameters within 7.54 seconds with the {\em slow} version where 
the {\em sensitivity} is  calculated,  
and within 0.068  seconds with the {\em fast} version where 
the {\em overlap complexity} is maximized
without considering the sensitivity explicitly.
The slow version of {\em rasbhari} is thus around 480 times faster 
than the exhaustive procedure in {\em CLARK-S}, while the fast version
is around $52,000$ times faster.  
The {\em theoretical} sensitivity of the three pattern sets is 0.999771 
for the default patterns from {\em CLARK-S}, 0.999811 for the 
{\em rasbhari} patterns with minimized overlap complexity  
and 0.999822 for the {\em rasbhari} patterns with maximized sensitivity.  

\begin{table}[h!]
\caption{
Read classification with {\em CLARK-S} \cite{oun:lon:15} with 
the default pattern set of the program and with 
the pattern set from 
{\em rasbhari} for the same parameter values, namely $n=3$ patterns of
length $\ell = 31$ and 
weight $w=21$. {\em Precision} and {\em sensitivity} of the 
classification are  reported at the
{\em species level} for five data sets from the literature.
\label{clark_tab}}
\vspace{2mm}
\begin{tabular}{|l|r|c|c|c|c|}
\hline
        &           & \multicolumn{2}{|c|}{Default pattern set} & \multicolumn{2}{|c|}{rasbhari} \\
\hline
Dataset & $\#$reads & Precision & Sensitivity & Precision & Sensitivity \\
\hline
HC1     & 999,998   &     97.69 &     90.36   & 97.69     & \bf 90.44  \\
HC2     & 999,991   &     96.45 &     88.11   & 96.45     & \bf 88.18  \\
simHC   & 116,771   &     97.20 &     90.53   & 97.20     & \bf 90.54  \\
simMC   & 97,495    & \bf 98.75 &     95.09   & 98.73     &     95.09  \\
simLC   & 114,457   & \bf 98.29 & \bf 94.26   & 98.28     &     94.25  \\
\hline
\end{tabular}
\end{table}

To evaluate the classification accuracy of {\em CLARK-S} with these three 
pattern sets experimentally, 
we used five data sets from the literature, namely two sets,
{\em HC1} and {\em HC2}, from the {\em MetaPhlAn} project
\cite{seg:wal:bal:etal:12}
and three sets, {\em simHC, simMC and simLC}, from the {\em  FAMeS}  databases 
\cite{mav:iva:bar:etal:07}.  For each of these data sets, we calculated 
{\em precision} and {\em sensitivity} 
of the classification at the {\em species level} 
as defined in \cite{oun:wan:clo:lon:15}. 
That is, for a classification task where objects are to be assigned to
classes,   
  {\em precision} is defined as the fraction of correct assignments
among the total number of assignments, while {\em sensitivity} is
the ratio between the number of correct assignments and
the number of objects to be classified. 
The two values are not the same since not every object is necessarily
assigned to one of the classes; {\em precision} is always larger than 
or equal to {\em sensitivity} since  the denominator in 
the definition of precision is smaller or equal to the denominator in the
definition of sensitivity. 
Since this definition of {\em sensitivity} refers to the ability of
a program to correctly classify objects, it is not to be confused with
the sensitivity in database searching as discussed above.  
Table \ref{clark_tab} summarizes precision and sensitivity of {\em CLARK-S}
with its default pattern set and with a pattern set generated by
{\em rasbhari}.

\begin{figure}[h!]
\begin{center} 
\includegraphics[width=6cm]{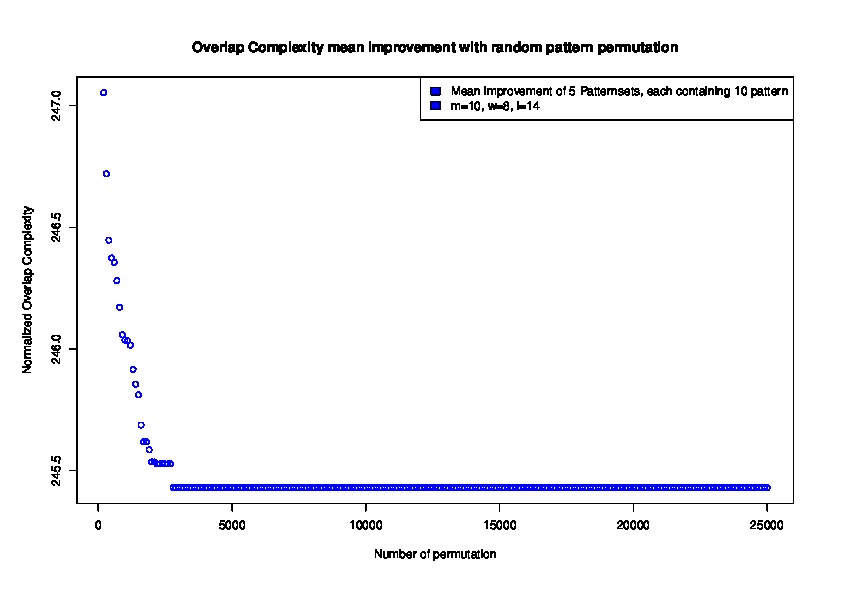}
\includegraphics[width=6cm]{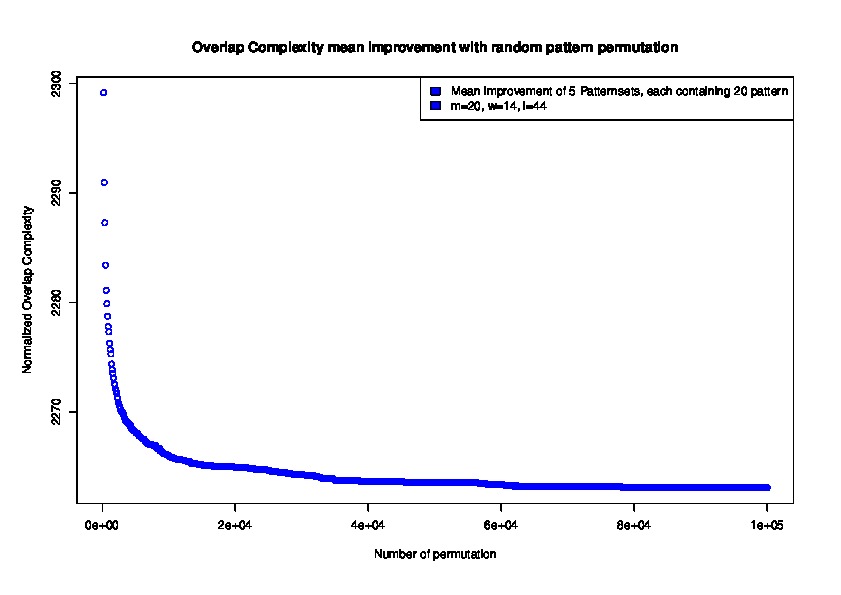}
\includegraphics[width=6cm]{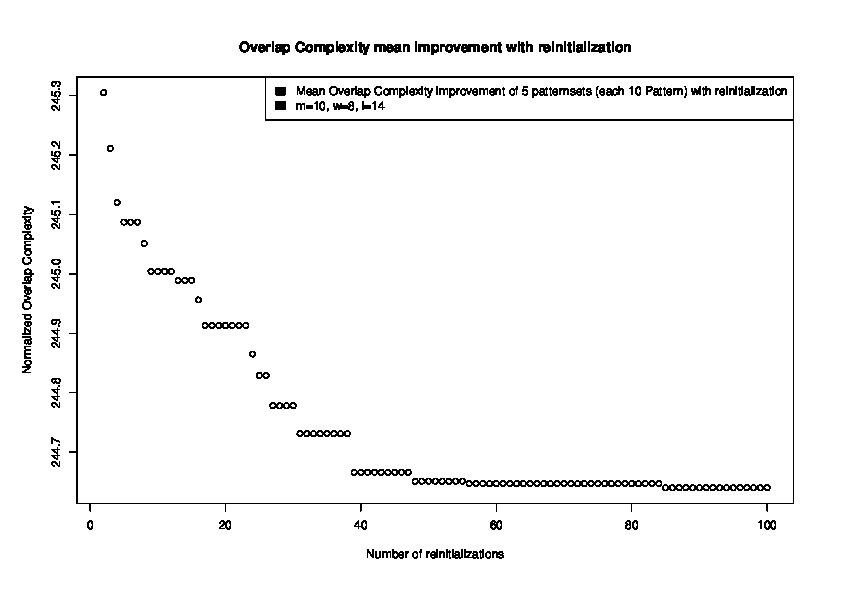}
\end{center}
\caption{Normalized overlap complexity {\em (OC)} of pattern sets depending 
on the number of iteration steps in our algorithm. The first two
plots show how the {\em OC}  is reduced in a single round of the  
hill-climbing algorithm for different parameters. 
For a set of $m=10$ patterns  of length $\ell =14$ and weight $w=8$, the
algorithm converges after around 3,000 iteration steps of hill-climbing
(upper plot); for a set of $m=20$ patterns of length $\ell = 44$ and
weigth $w=14$, it converges after around 80,000 steps (middle
plot).  The lower plot shows how the {\em OC} is improved if the
hill-climbing algorithm is run multiple times and the best result
of all runs is returned. 
\label{fig_opti}
}
\end{figure}

\subsection*{Improvement of {\em OC} by hill climbing}

Figure \ref{fig_opti} shows how the overlap complexity {\em (OC)} 
of pattern sets produced by {\em rasbhari} depends on the
number of iteration steps carried out in the hill-climbing algorithm. 
For a set of $m=10$ patterns of length $\ell=14$ and weight $w=8$, a single
run of the hill-climbing procedure converges after around 3,000 steps; for 
$m=20, \ell=44, w=14$, it converges after around 80,000 steps. 
The {\em OC} is further improved if the hill-climbing procedure is 
run multiple times and the best result of these runs is used. 

\section*{Discussion}
We developed a program called {\em rasbhari} to calculate sets of binary
{\em patterns} -- or {\em spaced seeds}, as they are often called -- 
 for read mapping, database searching and alignment-free sequence comparison. 
For sequence-homology searching, {\em rasbhari} optimizes  
the {\em sensitivity} of pattern sets, {\em i.e.} the probability of obtaining
at least one hit between a query and a database sequence that share 
a gap-free homology of a given length and with a given match
probability between nucleotides.  
Since the sensitivity of a pattern set is expensive to calculate,
our algorithm optimizes the {\em overlap complexity} of the produced pattern 
sets which is closely related to its sensitivity. 
We use a {\em hill-climbing} algorithm, similar to the one used in {\em SpEED},
to minimize the  overlap complexity. Unlike {\em SpEED}, however, our algorithm
does not calculate the overlap complexity of {\em all} neighbours of a current
pattern set, but modifies those patterns first that contribute most to the
overlap complexity of the current pattern set. If maximizing the 
{\em   sensitivity} in database searching, we calculate the sensitivity
of the current pattern set after a certain number of iterations and, finally, 
the pattern set with the overall highest sensitivity is returned.  

As a fast alternative, {\em rasbhari} can minimize the overlap complexity 
alone,  without calculating the sensitivity of pattern sets. 
This option is useful in situations where 
large pattern sets are needed for which it would take too long   
to calculate the sensitivity. 
As a third option,  {\em rasbhari} can minimize the variance of the number
$N$ of spaced-word matches in alignment-free sequence comparison which 
is used by various methods 
to estimate phylogenetic distances between sequences.  
We could show that, mathematically, the variance of $N$ has a similar form
as the overlap complexity of a pattern set, so the same optimization 
algorithm can be used to minimize both of them.


In both homology searching and read classification, pattern sets generated  
by {\em rasbhari} are more {\em sensitive} than the default
pattern sets, so  
more homologies can be detected and more reads can be correctly
classified. At first glance,   
the increase in sensitivity that we obtained 
seems moderate; as shown in Table~\ref{sensi_tab}, the improvement  
is usually in the first or second digit after the decimal mark.
In database searching and read mapping, however, even small improvements
in sensitivity can lead to a large number of additional hits.   
Moreover, as these additional hits will be  mostly in the `twilight zone'
of low sequence similarity, they may be of particular interest to the user. 

In the context of read alignment, 
Ilie {\em et al.} pointed out that, with a $100$-fold coverage of the 
human genome, 
a 1 percent improvement in pattern sensitivity would mean that 
3 billion more nucleotides could be mapped \cite{ili:ili:big:11}, 
so the improvement that we achieved with {\em rasbhari} would 
still lead to tens or hundreds of millions of additionally mapped 
nucleotides.
In database searching, the situation is similar. 
If we consider, for example, homology regions of
length $H=64$ with a match probability of $p=0.8$ at the nucleotide level,
then with $w=11$, the sensitivity of {\em rasbhari} is improved
by less than $0.01$ percentage points compared to {\em SpEED},  
see Table~\ref{sensi_tab}. 
Note, however, that these sensitivity values are already 
close to $100\%$, so
the fraction of homologies that are {\em not} detected can be considerably  
reduced with the sligh improvement in sensitivity obtained with 
{\em rasbhari}.  
In our example, the number of homologies that are {\em missed} is reduced
by $>7 \%$ if {\em rasbhari} is used instead of {\em SpEED}.
With the same parameters, but with $p=0.7$, the sensitivity of both programs
is around $93 \%$. Here, the number of missed homologies is still reduced
by $3\%$ with {\em rasbhari}, compared to {\em SpEED}.

For alignment-free sequence comparison, pattern sets produced by 
{\em rasbhari} lead to more accurate phylogenetic distances than the
random pattern  sets that we previously used. While this result may not
be surprising, {\em rasbhari} is, to our knowledge, the first program
that  has been designed for this purpose and that can 
minimize the variance of the number of spaced-word matches.  
We therefore integrated {\em rasbhari} into 
our web server for alignment-free sequence comparison  \cite{hor:lin:bod:etal:14}.   

In read classification, the sensitivity of {\em CLARK-S} 
could be increased by $0.08$  and $0,07$ percentage points, respectively,
 for the largest data sets that we used,  {\em HC1} and {\em HC2}.
Each of these data sets contains around one million reads, so the improvement
in sensitivity that we achieved with {\em rasbhari} means that 
800 more reads from {\em HC1} and 700 more from {\em HC2}  
could be correctly classified by  {\em CLARK-S}. 
This is remarkable, since the classification accuracy of {\em CLARK-S} is
already very high, so it is hard to further improve the program. 
An interesting question in the context of {\em CLARK-S} is how the
length and weight of the patterns influence its accuracy. 
So far, it was difficult to investigate this question systematically, 
since the exhaustive method that the program uses by default,
is too time consuming.  
With the massive improvement in runtime that we achieved with 
{\em rasbhari}, it is now possible to systematically investigate how the 
accuracy of {\em CLARK-S} depends on the parameters of
the underlying pattern sets.

In the hill-climbing procedure, our default of 25,000 iteration steps 
was sufficient to obtain stable results for the parameter settings 
that we used in our benchmark studies; 
%
%
%
we were unable to further improve these results  
by increasing the number of iterations.  For different values of  
$m, w, \ell, p$ and $H$, however, it may be advisable to adapt the
number of iteration steps. 
If the number of patterns or their length and weight are 
increased, Figure~\ref{fig_opti} shows that 
a larger number of iteration steps can improve the results. 
The number of iterations within one round of hill climbing and the number of 
times the hill-climbing is carried out can be passed to 
{\em rasbhari} through 
the command line; the users can therefore adapt these 
parameter values for their particular applications if they 
do not want to use the default values of the program.

%

%
%
%




\section*{Author Contribution:}
LH conceived and implemented the algorithm, CL evaluated {\em rasbhari}
for phylogeny reconstruction,
RO and SL evaluated the accuracy of read classification with {\em CLARK-S},
BM guided the study and wrote the manuscript. All authors read and
approved the manuscript.

\section*{Acknowledgements}
We would like to thank Laurent No{\'e} for helpful discussions and 
for pointing out the similarity between the overlap complexity and 
the variance of the number of spaced-word matches. 
Lucian Ilie made useful comments on a previous version of this paper.

\bibliographystyle{abbrv} 

\bibliography{/home/bmorgen/my_papers/bibtex/all_papers}

\end{document}